\documentclass[]{aa}  
\usepackage{amsmath}
\usepackage{amssymb}
\usepackage{natbib}
\usepackage{graphicx}
\usepackage{txfonts}
\usepackage[english]{babel}
\usepackage{hyperref}
\usepackage{units}
\begin{document}
\title{Random sampling technique for ultra-fast computations of molecular opacities for exoplanet atmospheres}

\author{
	M. Min\inst{1,2}
}

\offprints{M. Min, \email{M.Min@sron.nl}}

\institute{
SRON Netherlands Institute for Space Research, Sorbonnelaan 2, 3584 CA Utrecht, The Netherlands
	\and
Astronomical institute Anton Pannekoek, University of Amsterdam, Science Park 904, 1098 XH, Amsterdam, The Netherlands
}

   \date{Last edit: \today}

 
  \abstract
   {Opacities of molecules in exoplanet atmospheres rely on increasingly detailed line-lists for these molecules. The line lists available today contain for many species up to several billions of lines. Computation of the spectral line profile created by pressure and temperature broadening, the Voigt profile, of all of these lines is becoming a computational challenge.}
   {We aim to create a method to compute the Voigt profile in a way that automatically focusses the computation time into the strongest lines, while still maintaining the continuum contribution of the high number of weaker lines.}
   {Here, we outline a statistical line sampling technique that samples the Voigt profile quickly and with high accuracy. The number of samples is adjusted to the strength of the line and the local spectral line density. This automatically provides high accuracy line shapes for strong lines or lines that are spectrally isolated. The line sampling technique automatically preserves the integrated line opacity for all lines, thereby also providing the continuum opacity created by the large number of weak lines at very low computational cost.}
   {The line sampling technique is tested for accuracy when computing line spectra and correlated-k tables. Extremely fast computations ($\sim3.5\cdot10^5$ lines per second per core on a standard current day desktop computer) with high accuracy ($\leq1$\% almost everywhere) are obtained. A detailed recipe on how to perform the computations is given.}
   {}


   \maketitle
%

\section{Introduction}

After an extremely successful decade of exoplanet discoveries, the exoplanet research field is rapidly moving into the age of characterisation. Spectroscopic observations of exoplanet atmospheres are becoming more and more detailed especially with the HST WFC3 as an important contributor. The launch of the James Web Space Telescope (JWST) next year will likely accelerate this development enormously. The increasing level of detail in the observations is accompanied by an increasing demand on the accuracy of the models trying to explain the data. A first step in modelling an exoplanet atmosphere is having the right opacities of both molecules and cloud particles. The hot atmospheres of many planets we can study today provide challenges for the databases of molecular lines created for the atmosphere of the Earth and solar system planets, like HITRAN \citep{2013JQSRT.130....4R} and HITEMP \citep{2010JQSRT.111.2139R}. To solve this problem the ExoMol project was initiated \citep{2012MNRAS.425...21T, 2016JMoSp.327...73T}. In this project a mixture is employed of first principles and empirically tuned quantum mechanical computations, providing the most complete list of transitions for a wide variety of molecules.

Having a line list that is as complete as possible is crucial for doing proper radiative transfer computations in high temperature atmospheric environments. However, computing the line opacities for a large number of lines can be computationally challenging. For example, the ExoMol line list of CH$_4$ contains on the order of $10^{10}$ lines \citep{2014MNRAS.440.1649Y}. Computing the exact pressure and temperature broadened Voigt profile for each of these lines is computationally extremely demanding. 
There are several approximate methods for computing Voigt profiles available in the literature \citep[e.g.][]{1982JQSRT..27..437H, Weideman1994, Zaghloul:2012:A9C:2049673.2049679}. Also, much effort is spend succesfully on making these approximate methods faster and more accurate \citep[see e.g.][]{Poppe:1990:MEC:77626.77629, 2007JQSRT.107..173L}. These methods all focus on obtaining a given accuracy of the exact shape of the Voigt profile by applying mathematical approximations to decrease the computation time. While these methods still require significant computation time, they are now routinely applied to compute Voigt profiles in many applications.
The fastest code able to compute Voigt profiles of large numbers of lines at this moment is the HELIOS-k code \citep{2015ApJ...808..182G}. This code is able to compute $\sim10^5$ lines per second on a dedicated NVIDIA K20 GPU based machine. This implies that the computation of $10^{10}$ lines still requires on the order of one day for a single point in pressure temperature space. Usually a grid of pressure and temperature points is required. Thus, there is the need for an even faster method. In addition, we have to make sure that there are no systematic errors in the computations because a small systematic error in a single line can become large when computed for $10^{10}$ lines. Thus, we seek a method that is statistically exact, preserves the integrated opacities and the average shape of the lines, and computes the line profile accurately for the stronger lines.

For most practical applications we are not interested in the detailed line shapes of all $10^{10}$ lines. Only a few of the strongest lines can be distinguished, while the weaker lines provide continuum opacity. However, the detailed properties of the line wings due to pressure broadening are very important to take into account \citep{2016MNRAS.458.1427H}. \citet{2017arXiv170605724Y} use these arguments to separate the linelist into a list of weaker lines, for which a pseudo-continuum is computed, and a list of stronger lines, for which the full line profiles are computed. The authors of that paper show that the number of strong lines is orders of magnitude smaller than the total number of lines, so the computation time required is reduced with a similar fraction. This is a very strong method, with the only drawback that a split into strong and weak lines is required. In this paper we take a different approach which automatically transitions from strong to weak lines.

The extremely large number of lines is ideal for statistical methods. In this paper these properties are used to construct a line sampling technique where the line shape is randomly sampled. The technique automatically focusses the computation time in constructing accurate line shapes for the strongest lines, while for the weakest lines the focus lies on creating the right continuum opacity. The statistical properties of the line shapes are computed without approximation, taking into account the detailed far line wings due to pressure broadening of the vey large number of lines. This creates accurately the expected continuum from these large number of line wings. The computation time of the line sampling technique scales favourably with the number of lines, meaning that for large line numbers increasing the number of lines has very little influence on the computation time required.

In section \ref{sec:opacities} the line sampling technique is presented in detail and the numerical implementation is explained. The method is applied to computing the CH$_4$ opacities and correlated-k tables in section \ref{sec:example}. Finally, the results are summarised in section \ref{sec:summary}.

\section{Opacity computations}
\label{sec:opacities}

\subsection{Statistical line profile sampling}

The large line lists available for species like CH$_4$ and H$_2$O contain so many lines that in low resolution spectra there are $10^5$ to $10^6$ lines per spectral resolution element. Thus, we are not so much interested in the exact line shapes as we are in the statistical distribution of the opacities. The correlated-k method simply samples the probability opacity distribution within a resolution bin. An efficient method for computing the correlated-k tables therefore focusses on computing these statistical properties, rather than the exact line shape of each separate line.

A molecular line is broadened by Doppler- and pressure broadening. The convolution of a Gaussian (Doppler) and Lorentz (pressure) profile creates the Voigt profile. Computing the exact shape of this convolution can be computationally demanding. Here we do not compute the convolution directly but sample the exact shape of the convolution by sampling the  Gaussian and a Lorentz profile. Sampling a probability distribution can be done by creating a number of $N$ samples of the frequency shift with respect to the line centre, and adding a fraction of $1/N$ of the integrated line to the frequency bin corresponding to that frequency shift. In the limit of $N\rightarrow\infty$ this gives the exact line shape. Standard algorithms exist for sampling probability distributions of various shapes. For an explanation on sampling of the deviates of a probability distribution the reader is referred to chapter 7.2 of \cite{1992nrfa.book.....P}. While no analytic expression exists for sampling the deviate of the Voigt probability distribution, we can construct the deviate exactly by sampling the deviates of the Gaussian and Lorentzian distributions of which the Voigt profile is a convolution.

The broadening of the line due to thermal velocity of the gas is given by the Gaussian probability distribution
\begin{equation}
\phi_\mathrm{therm}= \frac{1}{\sigma\sqrt{2\pi}}\exp\left\{-\frac{(\nu-\nu_0)^2}{2\sigma^2}\right\}
\end{equation}
where $\nu$ is the frequency of the radiation, $\nu_0$ the centre of the line, and $\sigma$ the thermal broadening parameter given by
\begin{equation}
\sigma=\frac{\nu_0}{c}\sqrt{\frac{2\,k_b\,T}{m_\mathrm{mol}}}
\end{equation}
with $k_b$ the Boltzmann constant, $T$ the temperature of the gas, $m_\mathrm{mol}$ the mass of the molecule, and $c$ the speed of light. 

The broadening of the line due to pressure effects is given by the Lorentz probability distribution
\begin{equation}
\phi_\mathrm{press}=\frac{\gamma/\pi}{\gamma^2+(\nu-\nu_0)^2},
\end{equation}
where $\gamma=\sqrt{\sum\gamma_i^2}$ with $\gamma_i$ is the broadening parameter due to pressure broadening of species $i$. The dependency of $\gamma_i$ on pressure and temperature is given by
\begin{equation}
\gamma_i=\gamma_{i,0} \frac{p_i}{P_0}\left(\frac{T_0}{T}\right)^n_i,
\end{equation}
where $\gamma_{i,0}$ is the broadening parameter at temperature $T_0$ and pressure $P_0$, $n_i$ is the pressure broadening exponent due to broadening by species $i$, and $p_i$ is the partial pressure of species $i$ ($p_i=f_i\cdot P$, with $f_i$ the abundance of species $i$ and $P$ the total pressure). In the ExoMol database there is broadening data available for H$_2$ and He, which are the most important collisional partners in the broadening of the lines \citep{2017JPhCS.810a2010Y}.

A Gaussian profile can be easily sampled using standard algorithms. Given random numbers $-1<\eta_1, \eta_2<1$,
\begin{equation}
\Delta\nu_\mathrm{therm}=\sigma\,\eta_1\sqrt{-2\frac{\ln\left(\eta_1^2+\eta_2^2\right)}{\eta_1^2+\eta_2^2}},
\end{equation}
samples the Gaussian profile.

Also a Lorentz profile is easily sampled. Given a random number, $-1<\eta<1$,
\begin{equation}
\Delta\nu_\mathrm{press}=\gamma\tan\left(\frac{\pi}{2}\eta\right),
\end{equation}
samples the Lorentzian profile.

\begin{figure}[!tp]
\centerline{\resizebox{\hsize}{!}{\includegraphics{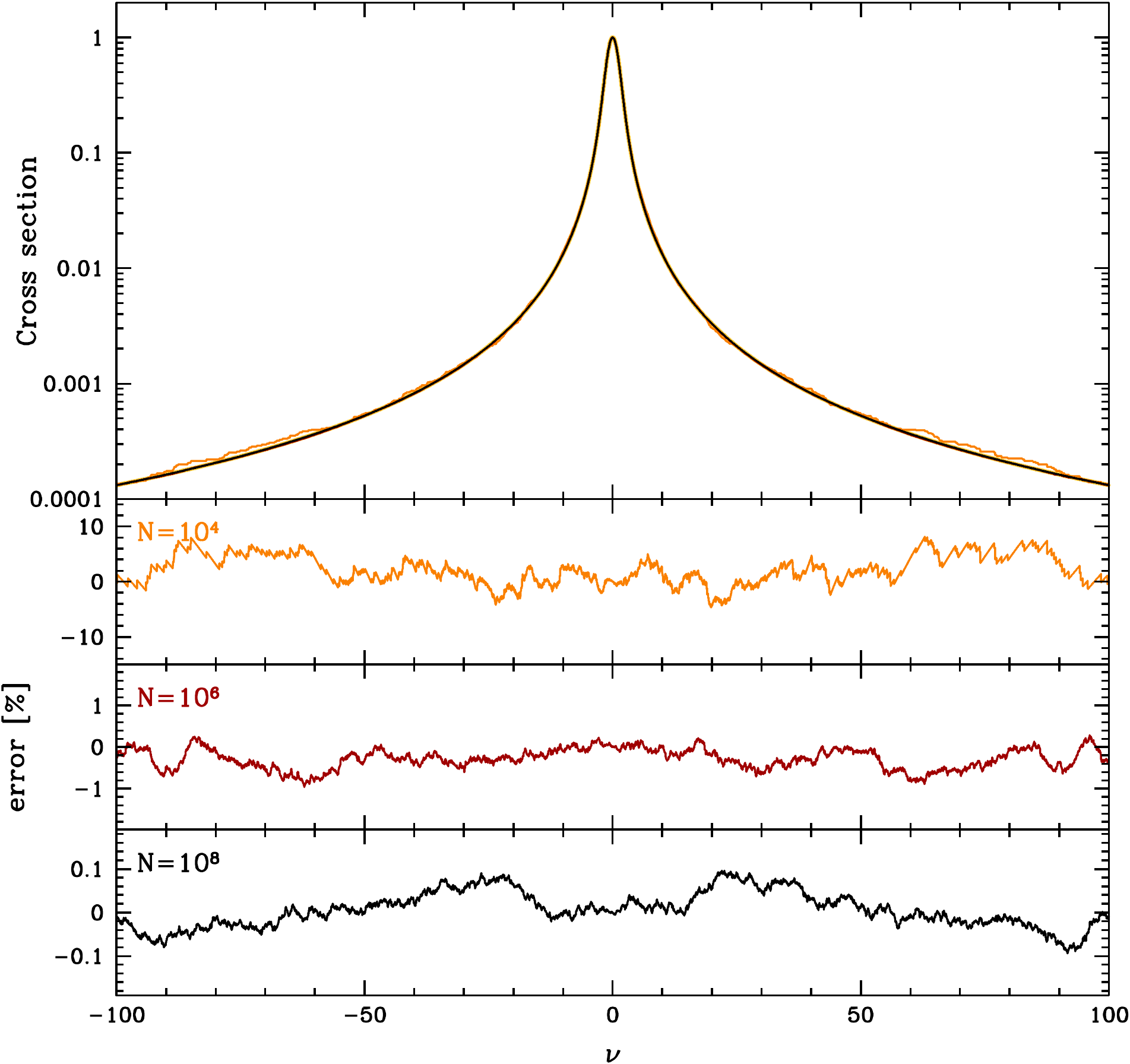}}}
\caption{Example Voigt profile sampled using different values for $N$. The three lower panels represent the relative error (in \%) with respect to a Voigt profile computed using a classical method. The frequency is in arbitrary units, $\gamma=\sigma=1$.}
\label{fig:Voigt}
\end{figure}

Both thermal and pressure broadening move energy of the line away from the line centre. The probability distributions give the fraction of the energy that is moved away from the line centre by a shift $\Delta\nu=\nu-\nu_0$. The combined effect of thermal and pressure broadening, the Voigt profile, is thus computed by the convolution of these two probability distributions. In other words, we move the energy away from the line centre by two shifting mechanisms implying that we are able to sample the Voigt distribution by
\begin{equation}
\Delta\nu=\Delta\nu_\mathrm{therm}+\Delta\nu_\mathrm{press}.
\end{equation}
We stress that by adding the frequency shifts in the sampling procedure, we compute the exact shape of the convolution of the two profiles and thus the exact shape of the Voigt profile. This procedure should not be confused with the pseudo-Voigt profile \citep[which introduces a small error at the advent of decreased computation time][]{Ida:nt0146}. In principle, the procedure to construct a Voigt profile is to use this sampling procedure to distribute the integrated opacity of each line over frequency space. We create $N$ random values for $\Delta\nu$ and add a fraction of $1/N$ of the integrated line opacity in the frequency bin corresponding to the frequency $\nu_0+\Delta\nu$, with $\nu_0$ the line centre. The disadvantage of this direct approach is that we need many samples to make sure no bins in the line wings are missed and end up with zero opacity. A better procedure is to spread the opacity over multiple bins. We spread the opacity of a sample over all bins in the frequency range $\nu_1\leq\nu\leq\nu_2$ given by,
\begin{equation}
\nu_1=\nu_0-|\Delta\nu_\mathrm{press}|+\Delta\nu_\mathrm{therm},
\end{equation}
\begin{equation}
\nu_2=\nu_0+|\Delta\nu_\mathrm{press}|+\Delta\nu_\mathrm{therm}.
\end{equation}
To obtain the Voigt line profile from this we have to weight each sample with a factor
\begin{equation}
w=\frac{\Delta\nu_\mathrm{press}^2}{\Delta\nu_\mathrm{press}^2+\gamma^2}.
\end{equation}
Finally, the profile sampled in this way is normalised to the integrated line opacity.

Often we would like the pressure broadening profile to have a cutoff at a certain value. In case a cutoff is preferred which is a multiple of $\gamma$, we can simply limit the random number $\eta$ to alter the range of $\Delta\nu_\mathrm{press}$. In case an absolute cutoff in terms of $\Delta\nu$ is preferred, a simple check is required if the sampled $\Delta\nu$ is within the given range. If not, we just draw a new sample. An important aspect of the sampling procedure is that the integrated line opacity is always conserved, independent on if and where the line wings are cut.

Constructing a smooth line profile using the above sampling procedure can be efficient, but the real efficiency gain is when we realise that we do not need a perfectly smooth line shape for each of the $10^5$ lines in each frequency bin. We can adjust the number of samples $N$ for each line, taking into account:
\begin{itemize}
\item The local line density; more lines per frequency interval means we can reduce $N$ and still maintain a proper statistical sampling of the opacity distribution in the low resolution frequency bins.
\item The line strength; weak lines will only create continuum, allowing to decrease $N$. We note that the integrated opacity is always exactly fixed, independent of $N$. Thus for very weak lines, even $N=1$ will create the required continuum opacity.
\end{itemize}
After extensive empirical testing and optimising efficiency and accuracy we take,
\begin{equation}
\label{eq:numbersamples}
N=\frac{R_H\,S}{200\,N_L\,R_L\,S_\mathrm{aver}},
\end{equation}
where $R_H$ is the spectral resolution used for sampling of the lines, $R_L$ is the low resolution required for the correlated-k tables, $N_L$ is the local number of lines in the resolution element corresponding to $R_L$, $S_\mathrm{aver}$ is the average line strength in the low resolution frequency bin, and $S$ is the strength of the line under consideration. The value of $N$ is restricted to $1\leq N\leq10^7$.

Finally, the full computational gain of this method is obtained by storing a large number of values for $\Delta\nu_\mathrm{therm}/\sigma$ and $\Delta\nu_\mathrm{press}/\gamma$. We can construct the line samples for each line with given $\gamma$ and $\sigma$ by randomly sampling from these two arrays. This avoids having to compute the $\ln$ and $\tan$ function too often. For both we use an array of $10^8$ pre-sampled values to avoid any possibility of duplicate sampled lines. 

In Fig.~\ref{fig:Voigt} we show a Voigt profile with $\gamma=\sigma=1$ for different values of $N$. The lower three panels in this figure give the error in percentage with respect to a Voigt profile calculated using a classical method.

\begin{figure}[!tp]
\centerline{\resizebox{\hsize}{!}{\includegraphics{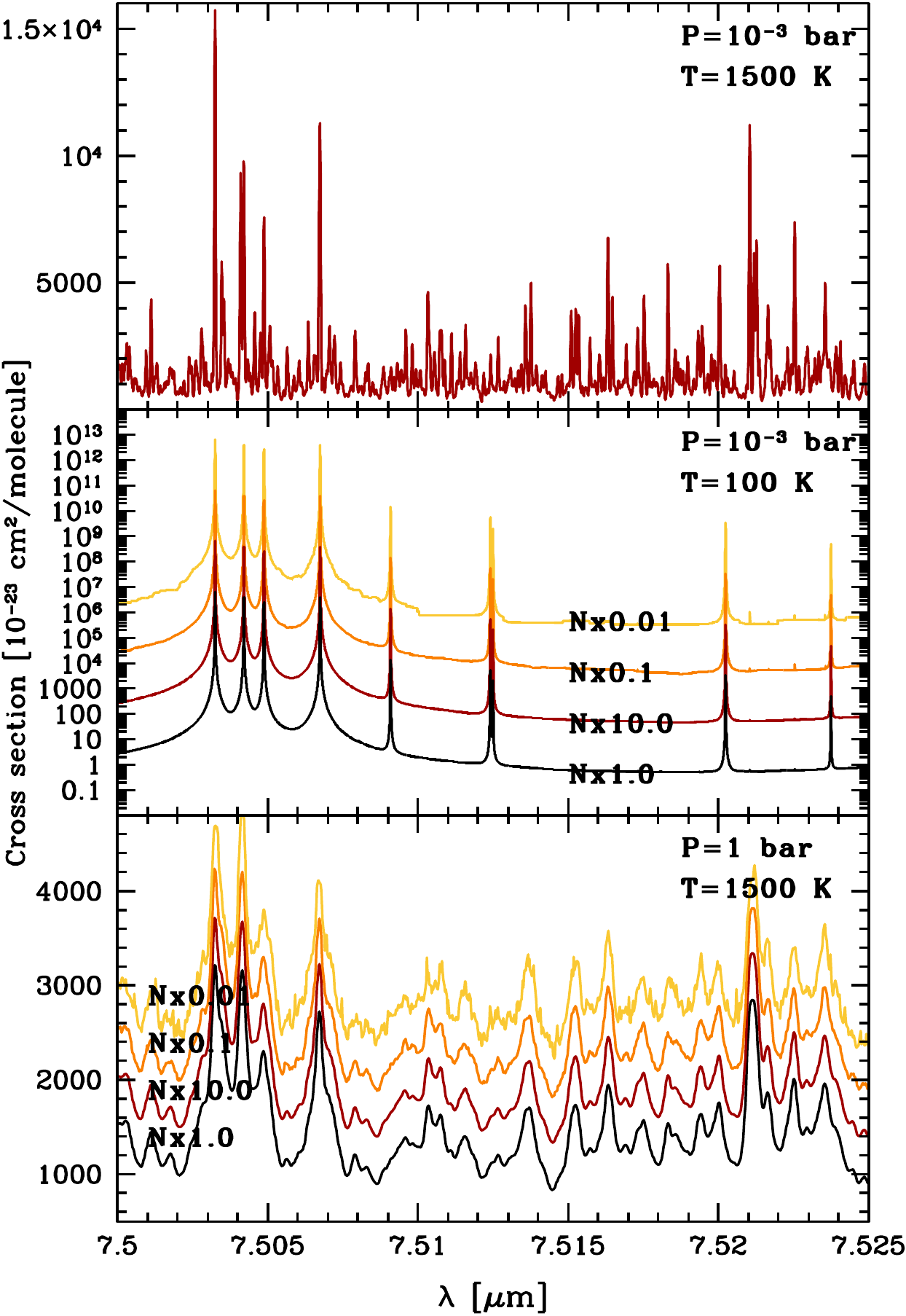}}}
\caption{Line spectra of methane for various pressures and tempartures computed using the line sampling technique. In the lower two panels also shown are the spectra computed with more or less samples. The spectra with different values for the number of samples are offset with respect to the $N\times1.0$ curves.}
\label{fig:CH4lines}
\end{figure}

\subsection{Correlated-k computations}

The correlated-k method for opacity sampling is a way to perform low resolution computations but still take into account the strong opacity fluctuations at high resolution. Correlated-k uses the assumption that within the low resolution frequency bin, radiative transfer computations with the same opacity give the same result. This means that we have to perform the radiative transfer computations for all opacities occurring within this frequency bin. The idea behind correlated-k method is to sort all opacities in a low frequency bin and transfer from the parameter $\nu$ (or $\lambda$) to a parameter $0<g<1$ that contains all opacities sorted. So the lowest opacities are at $g=0$ and the highest at $g=1$. The distribution of opacities is the same in $g$ or $\nu$, so the radiative transfer is the same when averaged over $\nu$ or over $g$. The big advantage is that the opacity curve in $g$ is much smoother (see examples in the next section), so we can sample the radiative transfer computations over just a few $g$ points, where the line opacities fluctuate enormously over $\nu$, requiring orders of magnitude more points.

It is not our aim here to discus detailed properties of the correlated-k method. For that we refer the reader to \citet{1989JQSRT..42..539G, 2015ApJ...808..182G, 2017A&A...598A..97A}.

\begin{figure}[!tp]
\centerline{\resizebox{\hsize}{!}{\includegraphics{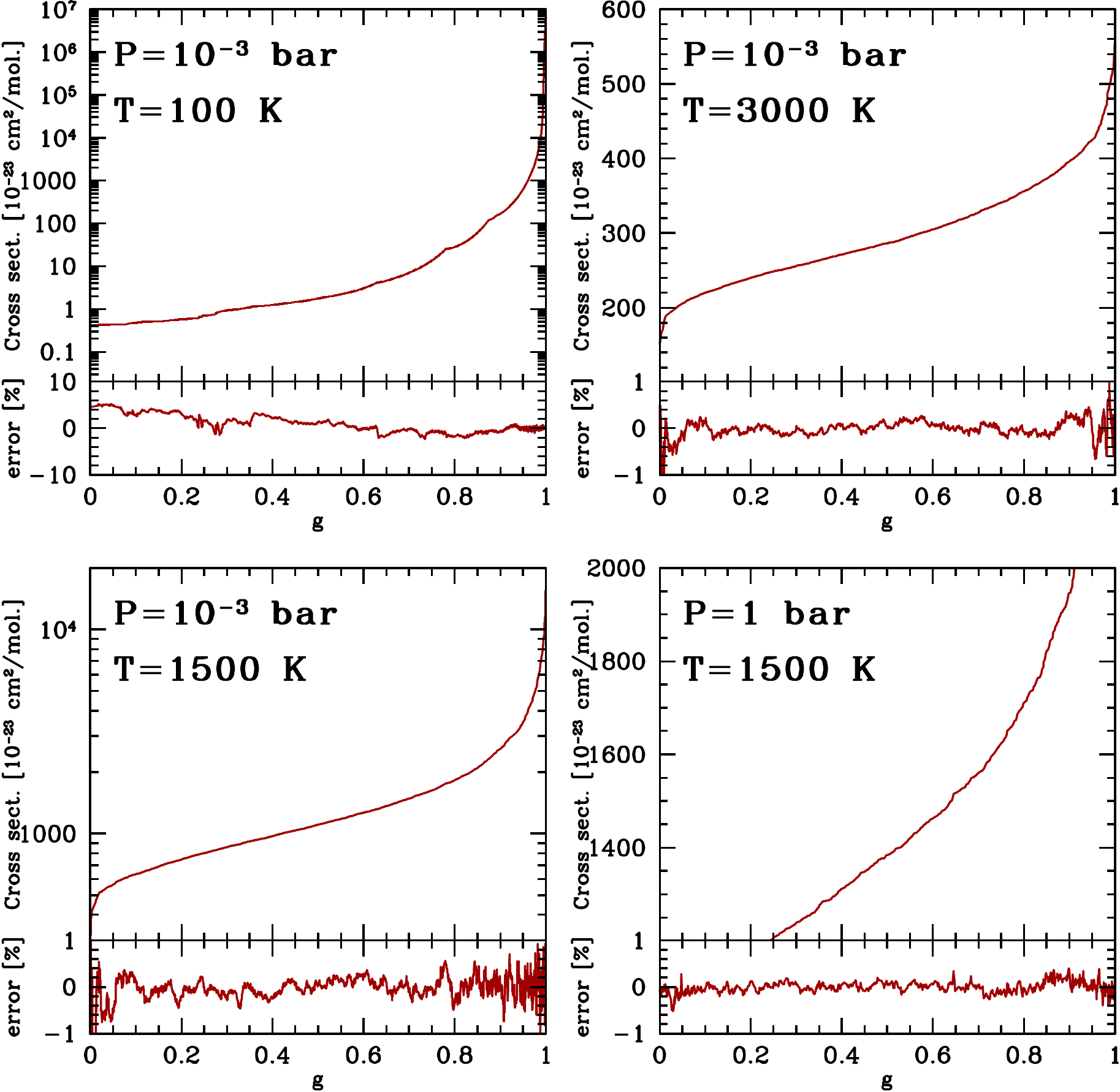}}}
\caption{Examples of correlated-k tables for methane for various pressures and temperatures. The k-tables correspond to a wavelength bin with $R=300$ around $\lambda=7.5\,\mu$m.}
\label{fig:CH4ktables8_0mic}
\end{figure}

\begin{figure}[!tp]
\centerline{\resizebox{\hsize}{!}{\includegraphics{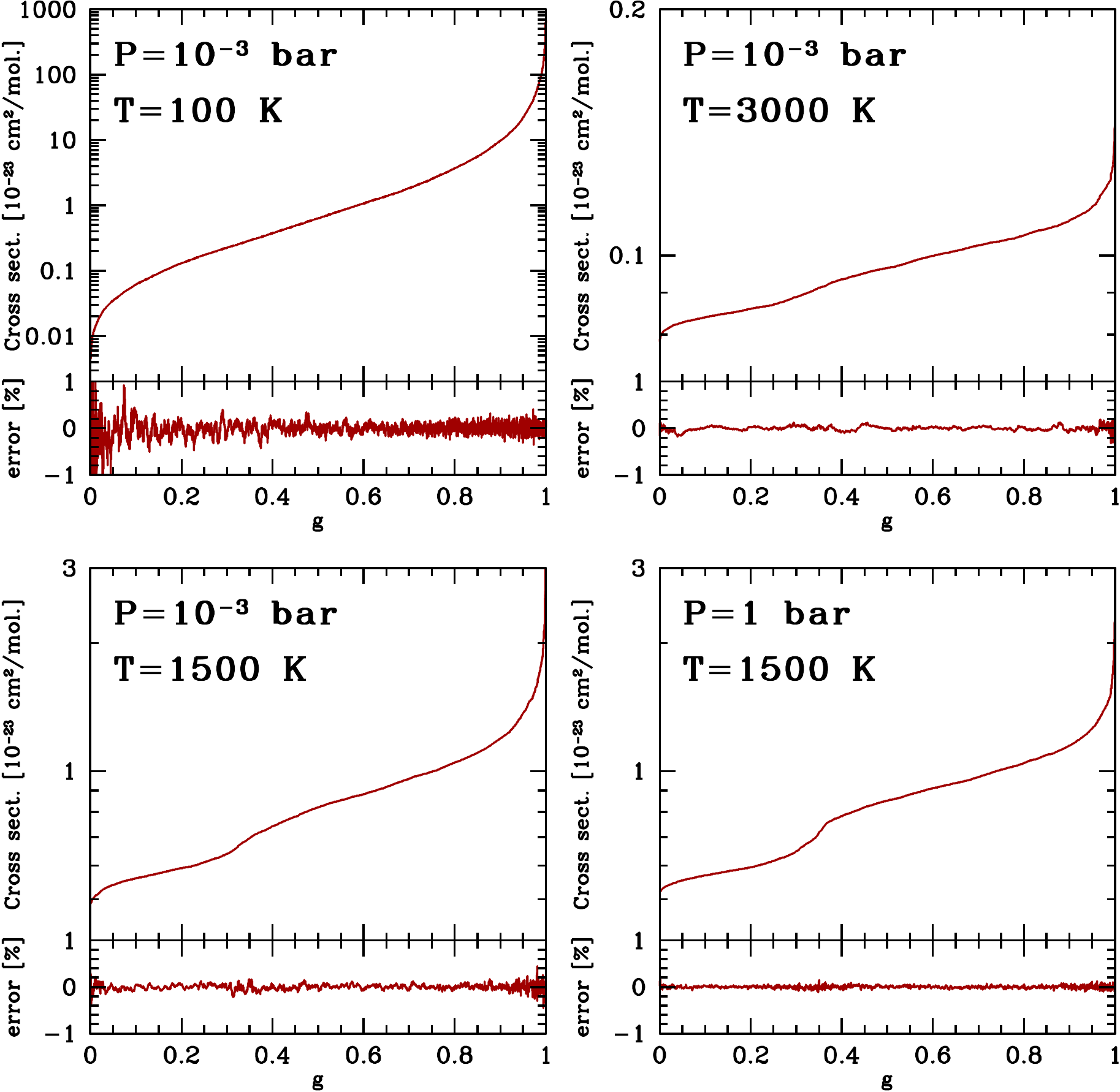}}}
\caption{Examples of correlated-k tables for methane for various pressures and temperatures. The k-tables correspond to a wavelength bin with $R=300$ around $\lambda=0.899\,\mu$m.}
\label{fig:CH4ktables0_9mic}
\end{figure}

\section{Example: CH$_4$}
\label{sec:example}

The line sampling method above is implemented in code reading in the files from the ExoMol database. The code produces line opacities and correlated-k tables\footnote{The computational code is available on a collaborative basis by contacting the author.}. As an example here we consider CH$_4$ in the near infrared \citep{2014MNRAS.440.1649Y}. In Fig.~\ref{fig:CH4lines} we plot a small part of the spectrum. Even this tiny wavelength range contains on the order of $10^6$ lines. The line profiles are sampled with a number of samples given by Eq.~\ref{eq:numbersamples} in the lowest curves. The other curves are for more or less samples. As can be seen, increasing the number of samples by a factor of ten does not significantly decrease the noise, while if we decrease the number of samples by ten or 100 we start to see sampling noise on the line opacities. In Fig.~\ref{fig:CH4ktables8_0mic} we plot the resulting correlated-k tables for a wavelength bin with $R=300$ around $\lambda=7.5\,\mu$m for different pressures and temperatures. We also plot here the relative difference with the correlated-k tables for ten times more samples. This gives an indication of the sampling error. As is expected, the error is larger for lower temperatures. This is caused by the fact that for lower temperatures the lines are narrower, and thus the sampling of the line wings is less accurate. In this case the wavelength bins relatively far from the line centre of the strong lines have poor statistics. However, we see that even in this case the error is below 10\% everywhere, while for all other cases it is even below 1\%. In Fig.~\ref{fig:CH4ktables0_9mic} we show the correlated-k tables for a wavelength bin around $0.9\,\mu$m. Also here the errors are below 1\% everywhere. This behaviour is representative for the entire spectrum from optical to the infrared. It is also found that the line spectra are more sensitive to noise than the correlated-k tables.

\begin{figure}[!tp]
\centerline{\resizebox{\hsize}{!}{\includegraphics{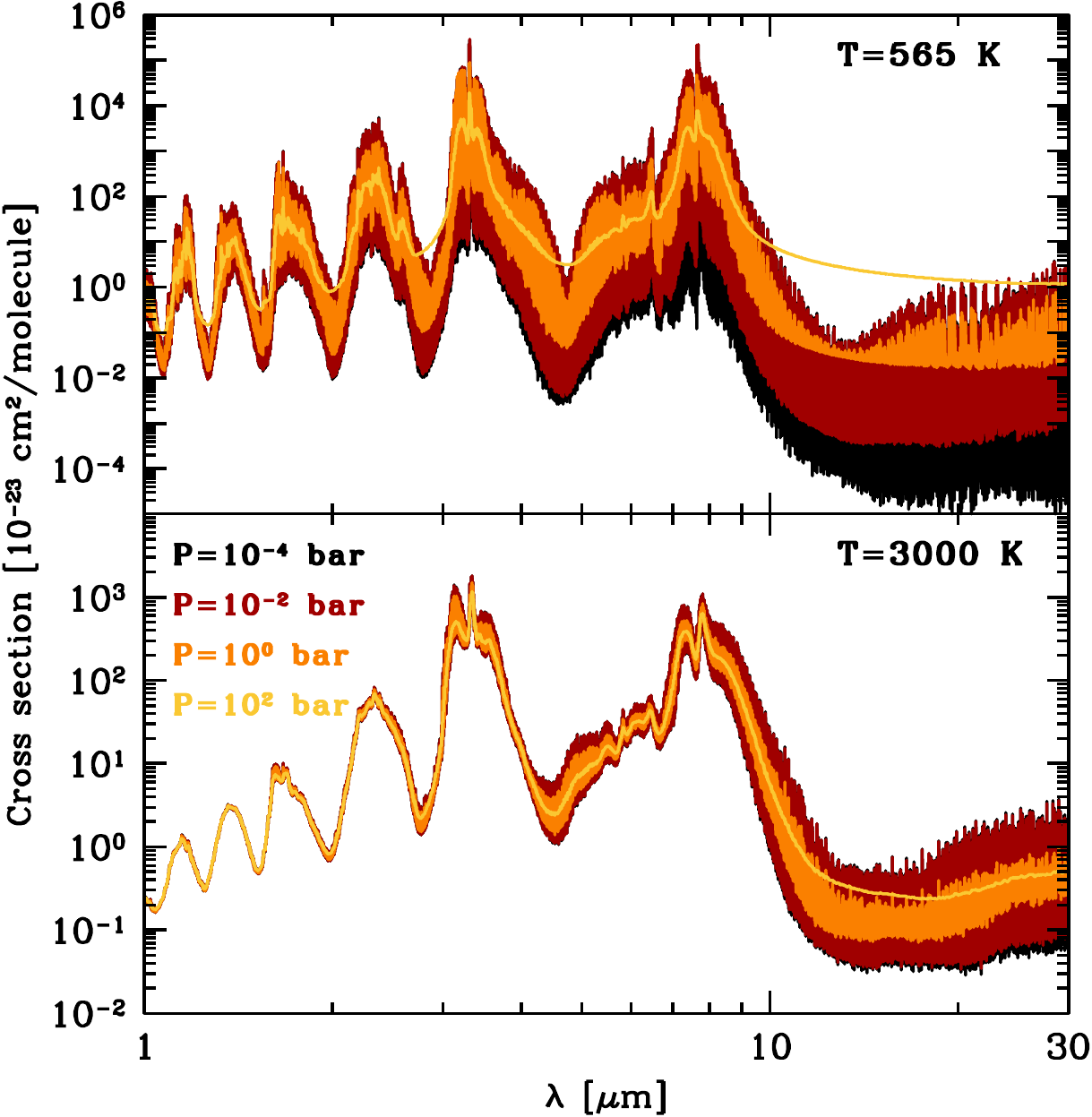}}}
\caption{The Methane absorption cross sections as a function of wavelength for various pressures and temperatures.}
\label{fig:CH4linesFull}
\end{figure}

The spectrum of CH$_4$ computed at two different temperatures and different pressures is shown in Fig.~\ref{fig:CH4linesFull}. We see that the opacity of the molecule converges towards a continuum for high pressures. This is caused by the very wide pressure broadened wings of the many lines. We note that for these computations we have considered the full Voigt profile without any wing cutoff. It is currently unknown where the pressure broadened line-wings should be cut off, but it can be seen that the effect of this is significant.

For methane at 1500\,K and 1 mbar the method described above computes around $3.5\cdot10^5$ lines per second for a single CPU on an iMac from 2011 (3.4 GHz Intel Core i7). So on a typical 8-core desktop machine, computing the full CH$_4$ spectrum with $10^{10}$ lines takes on the order of an hour and a half (this includes reading of the huge linelist files). The exact number can vary depending on the processor, molecule, temperature and pressure, but these numbers give a good indication of the speedup that the line sampling technique can provide for computing opacities from line lists. The code does not rely on dedicated machines, so can be run on arbitrary computer clusters. We note here also that by looking at Eq.~\ref{eq:numbersamples} we see that increasing the number of lines typically has only a small influence on the computation speed, since automatically the number of samples per line is reduced.

\section{Summary}
\label{sec:summary}

A method is provided for very fast computation of molecular opacities from large line lists. The method uses random sampling of the line profiles. There are no approximations of the shape of the lines or artificial line-cutoffs needed, the full Voigt profile is sampled for all lines. The method has the advantage that the computation time is not linearly proportional to the number of lines. When increasing the number of lines in the line list, we can decrease the number of samples per line to keep the computation time within reasonable limits.

It is shown that the method provides accurate correlated-k tables and even accurate high resolution line opacities. The broad, very far wings of the pressure broadened Voigt profile, creating continuum opacity for high pressures, can be fully computed with this method. The line sampling technique therefore provides a robust way of computing molecular opacities from large line list databases such as the ExoMol database for exoplanet opacities.

\begin{acknowledgements}
I would like to thank Ingo Waldmann for constructive and critical comments on an earlier version of this manuscript.\end{acknowledgements}

\bibliographystyle{aa}
\bibliography{biblio}

\end{document}